\documentclass[lettersize,journal]{IEEEtran}
\usepackage{amsmath,amsfonts,amssymb,amsthm}
\usepackage{dsfont} 
\usepackage{array}
\usepackage[caption=false,font=footnotesize,labelfont=rm,textfont=rm]{subfig}
\usepackage{textcomp}
\usepackage{stfloats}
\usepackage{url}
\usepackage{verbatim}
\usepackage{graphicx}
\def\BibTeX{{\rm B\kern-.05em{\sc i\kern-.025em b}\kern-.08em
    T\kern-.1667em\lower.7ex\hbox{E}\kern-.125emX}}
\usepackage{balance}
\usepackage{color}
\usepackage{cite}
\usepackage{algorithm}
\usepackage{algorithmicx}
\usepackage[noend]{algpseudocode}
\usepackage{booktabs}
\usepackage{diagbox}
\usepackage{multirow}
\theoremstyle{plain}

\usepackage{epstopdf}
\usepackage{orcidlink,hyperref}
\hypersetup{hidelinks,breaklinks=true}
\definecolor{blue}{rgb}{0, 0.45, 0.75}

\begin{document}
\title{Throughput Optimization for Multi-AP IEEE P802.11bq Networks Based on Combinatorial Multi-Armed Bandits}
\author{Anshan Yuan\hspace{0.5mm}\orcidlink{0009-0000-3293-5618}, Mingqi Han\hspace{0.5mm}\orcidlink{0000-0002-0061-7321}, Xinghua Sun\hspace{0.5mm}\orcidlink{0000-0003-0621-1469},~\IEEEmembership{Member,~IEEE}
\thanks{
Anshan Yuan, Mingqi Han, and Xinghua Sun are with the School of Electronics and Communication Engineering, Sun Yat-sen University, Shenzhen, China (e-mail: \url{yuanansh@mail2.sysu.edu.cn}; \url{hanmq@mail2.sysu.edu.cn}; \url{sunxinghua@mail.sysu.edu.cn}).
}
}

\maketitle

\begin{abstract}
IEEE P802.11bq targets integrated Wi-Fi operation in which sub-7GHz links support contention and control signaling, while millimeter-wave (mmWave) links deliver high-rate directional payloads. In dense multi-access point (AP) deployments, this architecture induces a coupled physical-layer (PHY)/medium-access-control (MAC) configuration problem: carrier-sensing and contention parameters determine channel access and Request-to-Send/Clear-to-Send (RTS/CTS) reliability, while beamwidth and modulation-and-coding scheme (MCS) margins govern mmWave interference, training overhead, link robustness, and rate. This paper addresses distributed throughput optimization for dense multi-AP IEEE P802.11bq networks. We develop a packet-level model that jointly captures cross-link carrier-sense multiple access with collision avoidance (CSMA/CA), sub-7GHz RTS/CTS exchange, beam-training overhead, directional mmWave interference, signal-to-interference-plus-noise-ratio (SINR)-based MCS selection, and retransmissions. The resulting configuration problem is formulated as a multi-group combinatorial multi-armed bandit (CMAB), where each AP selects its contention window, clear-channel assessment threshold, beamwidth, and MCS reservation margin from finite candidate sets. Inspired by combinatorial successive accept-reject methods, we propose a group-wise feasible CSAR variant that uses Hadamard-guided feasible exploration to estimate empirical ranking scores and eliminate low-performing candidates within each parameter group. Simulations show that the proposed scheme improves aggregate and per-AP throughput over the considered Thompson-sampling baseline across most AP densities and reduces throughput stabilization time by approximately 49\% under the evaluated settings. The learned configurations reveal that high throughput requires a balance among control-channel aggressiveness, mmWave spatial reuse, beam-training cost, and MCS robustness, rather than simply minimizing collisions or maximizing the PHY rate.
\end{abstract}


\section{Introduction}\label{sec:introduction}
\subsection{Research Background}
Wireless local area networks (WLANs) are evolving toward increasingly dense and capacity-intensive deployments. Emerging applications, including industrial digital twins, immersive communication, and interactive extended reality, require stable, high-throughput wireless connectivity over shared indoor spectrum~\cite{bnpar, survey_multi_ap_coordination}. These services are commonly deployed in office campuses, commercial complexes, and manufacturing facilities, where extensive coverage requirements and high user densities render single-access-point (AP) operation inadequate. Consequently, WLAN performance is increasingly influenced by interactions among neighboring APs rather than solely by the capabilities of an individual AP.

Millimeter-wave (mmWave) communications offers abundant spectrum and surpports highly directional transmissions, making it a promising technology for dense indoor WLAN deployments. However, mmWave links are particularly susceptible to blockage, severe path loss, beam misalignment, and rapidly varying interference. These characteristics make medium access control (MAC) significantly more challenging than in conventional sub-7GHz WLANs. A promising approach is integrated mmWave operation, in which robust sub-7GHz links support channel access and control signaling, while mmWave links provide high-rate directional data transmission \cite{bqpar}.

\subsection{Standardization Progress in IEEE P802.11bn and IEEE P802.11bq and Our Motivations}
IEEE 802.11be, also known as Wi-Fi~7, introduced multi-link operation at the MAC layer, enabling coordinated access across multiple links in the 2.4, 5, and 6~GHz bands~\cite{ieee80211be}. Building on this evolution, the IEEE 802.11 Working Group has initiated two complementary standardization efforts that align with the trends discussed above: IEEE P802.11bn for reliability-oriented dense WLAN, and IEEE P802.11bq for integrated mmWave data transmission.

IEEE P802.11bn, commonly associated with ultra high reliability, places greater emphasis on reliability, latency determinism, and robustness in dense and interference-prone WLAN~\cite{bnpar, survey_multi_ap_coordination}. Multi-AP operation is one important direction in this context, since overlapping basic service sets create coupled contention, hidden-node effects, and unstable interference conditions~\cite{galati2024wifi8}. While explicit AP-to-AP coordination can improve system-level performance, it also incurs signaling overhead, synchronization requirements, and implementation complexity in dense deployments. This observation motivates a complementary and lightweight approach based on autonomous transmission-parameter optimization, whereby each AP adapts its IEEE P802.11bq configuration using local observations while implicitly responding to the contention and interference generated by neighboring APs.

In parallel, earlier mmWave WLAN standards, such as IEEE 802.11ad \cite{ieee80211ad} and IEEE 802.11ay \cite{ieee80211ay}, demonstrated the feasibility of high-rate short-range communications. However, their physical layer (PHY)/MAC designs were only loosely integrated with mainstream sub-7GHz WLAN operation, limiting ecosystem interoperability and hindering large-scale commercial adoption~\cite{reshef2022future}. To address this limitation, IEEE P802.11bq was initiated to support non-standalone WLAN operation in the 42--71 GHz unlicensed bands while reusing sub-7GHz access and control mechanisms \cite{bqpar,liu2025wifi8}. In this architecture, the sub-7GHz link anchors carrier sensing, random access, and control signaling, whereas the mmWave link is employed for high-throughput directional payload transmission.

IEEE P802.11bq integrates reliable channel access with high bandwidth and enhanced spatial reuse. However, it also introduces a coupled cross-layer optimization problem. The sub-7GHz contention parameters determine when an AP obtains a transmission opportunity, while the beamwidth and modulation-and-coding scheme (MCS) selection in mmWave links determine whether that opportunity can be converted into successful payload delivery. A conservative contention policy may reduce collisions but underutilize mmWave spatial reuse; an aggressive policy may increase concurrent transmissions but cause Request to Send/Clear to Send (RTS/CTS) collisions, hidden-terminal effects, or packet losses due to insufficient SINR. Similarly, narrow beams improve link budget and interference suppression at the cost of beam-training overhead, while conservative MCS margins improve robustness at the cost of PHY rate.

These coupled effects make static or manually tuned configurations fragile in dense deployments. Moreover, the performance of a parameter tuple can only be observed after the complete packet service process, including contention, control exchange, beam training, data transmission, and possible retransmission. This motivates an online and distributed optimization framework that learns from local throughput feedback while preserving the grouped structure of IEEE P802.11bq PHY/MAC parameters.

\subsection{Contributions and Paper Organization}
The main contributions are summarized as follows:
\begin{itemize}
    \item We develop a multi-AP IEEE P802.11bq framework and formulate the associated distributed PHY/MAC configuration optimization problem. The model captures the coupled effects of sub-7GHz contention, RTS/CTS reservation, directional interference, SINR-based beamwidth and MCS selection, and retransmission. Based on this model, each AP jointly selects its contention window, clear-channel assessment (CCA) threshold, beamwidth, and MCS reservation margin to maximize locally observed throughput without requiring centralized coordination or explicit interference maps.

    \item We develop a CSAR-inspired group-wise feasible CMAB algorithm for IEEE P802.11bq configuration optimization. Building on existing combinatorial successive accept reject approach, the algorithm decomposes a complete PHY/MAC tuple into elementary arms, but adds a feasibility mapping so that exactly one candidate is executed from each parameter group. The resulting method combines Hadamard-guided exploration, empirical ranking, and group-wise pruning to reduce the exploration burden while learning from end-to-end throughput feedback.

    \item We conduct packet-level simulations to evaluate both throughput gains and learned parameter-selection behavior. The proposed scheme improves aggregate and per-AP throughput over the Thompson-sampling based IEEE P802.11bq baseline for most AP densities and avoids the severe degradation of fixed configurations. The results reveal a key insight: achieving high throughput requires balancing control-channel aggressiveness, mmWave spatial reuse, beam-training overhead, and MCS robustness, rather than merely minimizing collisions or consistently selecting the highest PHY rate.
\end{itemize}

The remainder of this paper is organized as follows. Section~\ref{sec:related work} reviews related studies on dense multi-AP WLAN and mmWave WLAN parameter optimization. Section~\ref{sec:System Model And Preliminary Analysis} presents the multi-AP IEEE P802.11bq system model, the cross-link carrier-sense multiple access with collision avoidance (CSMA/CA) procedure, and the throughput optimization objective. Section~\ref{sec:Algorithm and Optimization} introduces the CSAR-inspired group-wise feasible CMAB method and its integration with the packet-level access process. Section~\ref{sec:Simulation} reports the simulation results and analyzes the learned parameter-selection behavior. Finally, Section~\ref{sec:Conclusion_discussion} concludes the paper and discusses future research directions.

\section{Related Work}\label{sec:related work}
\subsection{Multi-AP WLAN Optimization}
Dense WLAN have motivated extensive research on multi-AP coordination (MAPC), spatial reuse, and distributed access optimization. In Wi-Fi 7 and beyond, MAPC has been studied as a way to organize cooperation among neighboring APs, with representative mechanisms including coordinated OFDMA, coordinated spatial reuse, coordinated beamforming, and joint transmission~\cite{survey_multi_ap_coordination}. These mechanisms show that inter-AP cooperation can improve spectrum utilization and service predictability in overlapping basic service sets. However, practical deployment remains challenging because joint decision-making, coordination signaling, and synchronization become increasingly costly as AP density grows~\cite{shafin2025p2p}.

This limitation has led to a complementary line of work on distributed or lightweight optimization in dense WLANs. A central issue is that neighboring APs often have heterogeneous carrier-sensing views: an AP may defer because of locally sensed energy, while another AP may create harmful interference at a receiver that is not visible from the transmitter side \cite{sun2026throughput}. Such mismatches between carrier sensing and actual interference lead to hidden-node effects, unpredictable collisions, and spatial-reuse tradeoffs. Related analytical studies of multi-link mmWave WLAN access also reveal this tension. For example, cross-link RTS/CTS mechanisms can improve protection by using sub-7GHz links for control exchange, but stronger protection may also reduce concurrent mmWave transmissions and hence network throughput~\cite{chen2025cross}.

Centralized multi-AP scheduling has been investigated as another solution, particularly for latency- or freshness-sensitive traffic. A software-defined WLAN controller can jointly schedule transmissions across APs to improve information freshness under inter-AP interference, and coordinated policies outperform distributed single-AP baselines~\cite{rajagopalan2026optimizing}. Nevertheless, such approaches usually require a controller, timely network-wide state information, and nontrivial computation. Learning-based mechanisms, including DRL, MARL, and lightweight bandit-style adaptation, have therefore been explored for carrier-sensing adjustment, spatial-reuse control, coordinated scheduling, and decentralized resource adaptation~\cite{huang2022dqn,du2025deep,wojnar2025HMAB,wojnar2025CSR,jung2026csrSharedAPIndex,zhang2024MARLCBF,wilhelmi2026decentralized}.

Overall, existing multi-AP WLAN studies establish that dense deployments must balance distributed contention, asymmetric local observations, interference uncertainty, and the overhead of explicit coordination. However, most of them focus on sub-7GHz WLANs or on coordination mechanisms themselves. They do not directly address the IEEE P802.11bq setting in which sub-7GHz contention decisions and mmWave data-plane parameters jointly determine throughput. This gap motivates the autonomous multi-AP optimization considered in this paper, where each AP learns from local throughput feedback rather than relying on centralized scheduling or explicit AP-to-AP coordination.

\subsection{MmWave WLAN Parameter Optimization}
The feasibility of mmWave WLAN is supported by the availability of wide spectrum and highly directional transmission, but it is constrained by severe path loss, blockage, deafness, beam-management overhead, and directional interference. General mmWave surveys have characterized these opportunities and challenges across the physical, MAC, network, and cross-layer levels~\cite{niu2015survey,wang2018comprehensive,he2021phySurvey}. A common conclusion is that mmWave networking cannot be optimized at the PHY layer alone, because beam directionality changes initial access, carrier sensing, scheduling, and interference control.

At the WLAN level, IEEE 802.11ad introduced directional multi-gigabit communication in the 60-GHz band with hybrid contention-based and scheduled access as well as beam-training procedures~\cite{ieee80211ad, nitsche2014ad}. IEEE 802.11ay further enhanced mmWave WLAN operation through channel bonding, MIMO transmission, improved channel access, spatial sharing, and beam tracking~\cite{ieee80211ay, ghasempour2017ay,zhou2018aySurvey}. Related regional and protocol developments, such as IEEE 802.11aj and dynamic bandwidth control, also investigated mmWave WLAN operation and coexistence issues~\cite{wang2014aj,chen2014dbcMAC}. These studies provide the foundation for high-rate mmWave Wi-Fi, but they largely treat mmWave operation as a relatively separate WLAN mode.

A major difficulty of standalone mmWave WLANs is that directional communication weakens conventional carrier sensing. Stations may fail to detect ongoing transmissions outside their current beam direction, causing deafness, hidden-terminal effects, and unstable contention behavior. Prior work has addressed these issues through directional MAC protocols, analytical MAC modeling, adaptive control exchanges, and hybrid resource allocation~\cite{shokriMACPerspective,chandra2017adMAC,akhtar2018DMBS}. Another important direction is multi-band operation, where a lower-frequency interface carries robust control signaling while the 60-GHz interface carries high-rate data. Such designs can improve throughput and fairness under deafness and have been extended to more advanced multi-band or full-duplex MAC protocols~\cite{simMultibandDeafness,alkhrijah2023MBFDMAC}. IEEE P802.11bq follows this broader idea in a standardized Wi-Fi evolution path by using sub-7GHz operation to support access and control for mmWave data transmission.

These works indicate that integrated mmWave WLAN performance depends on multiple interacting parameters, including contention behavior, carrier-sensing threshold, beam training, beamwidth, interference level, and MCS selection. Existing optimization methods for WLAN and mmWave systems include theoretical modeling \cite{sun2026throughput,TGCN_yas,dai2013unified}, reinforcement learning \cite{Zhenyu2022MARLjsac,Zhenyu2025MARLtmc}, and MAB-based adaptation~\cite{wojnar2025HMAB,hmq2025bandit}. In the broader combinatorial bandit literature, combinatorial successive accept-reject algorithms have been developed for pure exploration and full-bandit feedback, including variants that use Hadamard matrices to estimate elementary-arm rewards~\cite{chen2014combinatorial, rejwan2020top}. However, most existing WLAN studies optimize a single component, require explicit state or interference information, or treat a complete configuration as an indivisible action. The existing CSAR theory also does not directly apply to our packet-level WLAN setting, because the reward is non-additive, affected by neighboring APs, and observed only after a group-wise feasible mapping of the Hadamard row. They therefore do not fully exploit the grouped structure of IEEE P802.11bq PHY/MAC parameters, where one candidate must be selected from each of several discrete parameter sets.

In summary, prior work has established the value of mmWave Wi-Fi, directional MAC design, and multi-band control. What remains insufficiently studied is online cross-layer parameter learning in dense multi-AP IEEE P802.11bq networks, where sub-7GHz contention and mmWave data transmission are coupled through the complete packet service process. This paper addresses this gap by formulating the joint selection of contention window, CCA threshold, beamwidth, and MCS reservation margin as a multi-group CMAB problem and solving it using local throughput observations.

\section{System Model and Preliminary Analysis} \label{sec:System Model And Preliminary Analysis}

We consider a dense multi-AP IEEE P802.11bq WLAN, where several APs are deployed in the same indoor service area and operate over shared spectrum resources. Let $\mathcal{N}=\{1,\ldots,N\}$ denote the set of APs, and let $\mathcal{S}_i$ denote the set of STAs associated with AP $i$. Since neighboring basic service sets are not isolated, the transmission decision of one AP affects both the carrier-sensing state and the mmWave reception quality of other AP--STA links. Therefore, the considered system is a coupled multi-AP, multi-link Wi-Fi network rather than a collection of independent single-link parameter-optimization problems.

The IEEE P802.11bq access procedure is modeled through an integrated multi-link architecture. The sub-7GHz interface provides reliable carrier sensing and efficient short control-frame exchanges, while the 60 GHz mmWave interface supports high-rate directional payload transmissions. This architecture is well suited for dense deployments: the sub-7GHz control plane offers larger coverage and more reliable coordination, whereas the mmWave data plane enables spatial reuse through directional beams. However, these two planes are tightly coupled. A conservative control-channel access policy may reduce collisions but leave mmWave spatial reuse underutilized, whereas an aggressive access policy may increase hidden interference and data-frame collisions.

\begin{figure*}[t]
    \centering
    \includegraphics[width=1\textwidth]{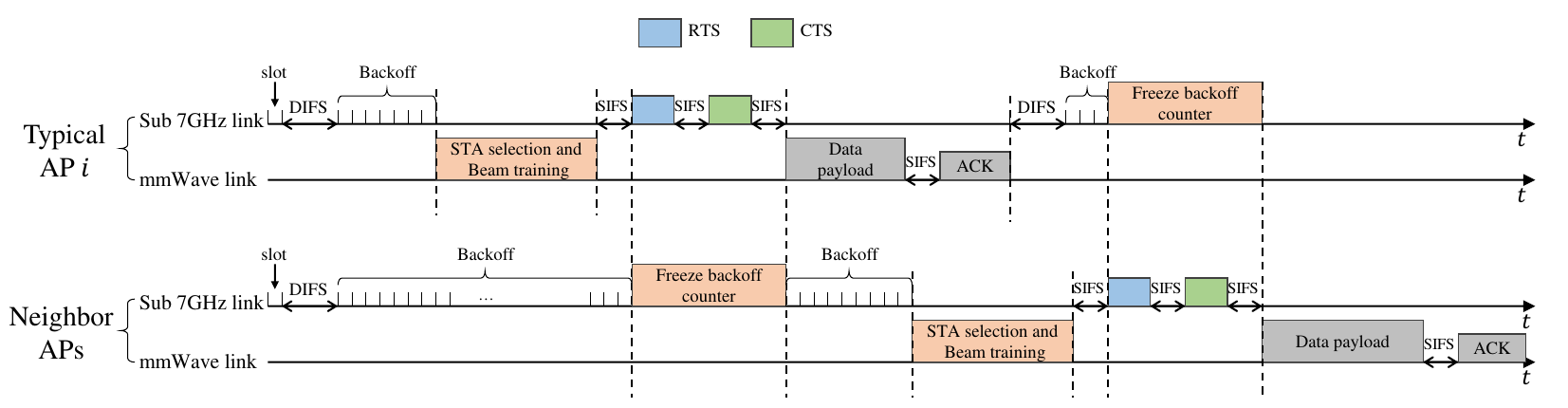}
    \caption{System model and the channel access procedure of the proposed multi-AP IEEE P802.11bq WLAN.}
    \label{fig: system_model}
\end{figure*}

\subsection{Cross-Link CSMA/CA Procedure}
Fig.~\ref{fig: system_model} illustrates the modeled multi-AP IEEE P802.11bq access process. For a typical AP $i$ and its neighboring APs, the sub-7GHz link carries carrier sensing, DIFS/SIFS timing, random backoff, and RTS/CTS control exchange, whereas the mmWave link is activated for STA selection, beam training, payload transmission, and ACK reception. When one AP occupies the sub-7GHz control channel, neighboring APs freeze their backoff counters and resume contention after the channel becomes idle. Once an RTS/CTS exchange succeeds, the reserved transmission opportunity is converted into a directional mmWave data transmission; if other APs obtain overlapping mmWave opportunities, their directional signals contribute to the interference seen by the intended STA. Thus, the figure highlights the key coupling captured by the model: sub-7GHz contention determines when an AP may transmit, while concurrent mmWave transmissions determine whether the selected beamwidth and MCS can sustain reliable payload delivery.

The packet-level operation of each AP is modeled as a continuous-time CSMA/CA process with RTS/CTS-assisted admission. As summarized in Fig.~\ref{fig: system_model}, the procedure consists of the following stages.

\subsubsection{Carrier sensing and backoff}
\begin{figure}[t]
    \centering
    \includegraphics[width=0.5\textwidth]{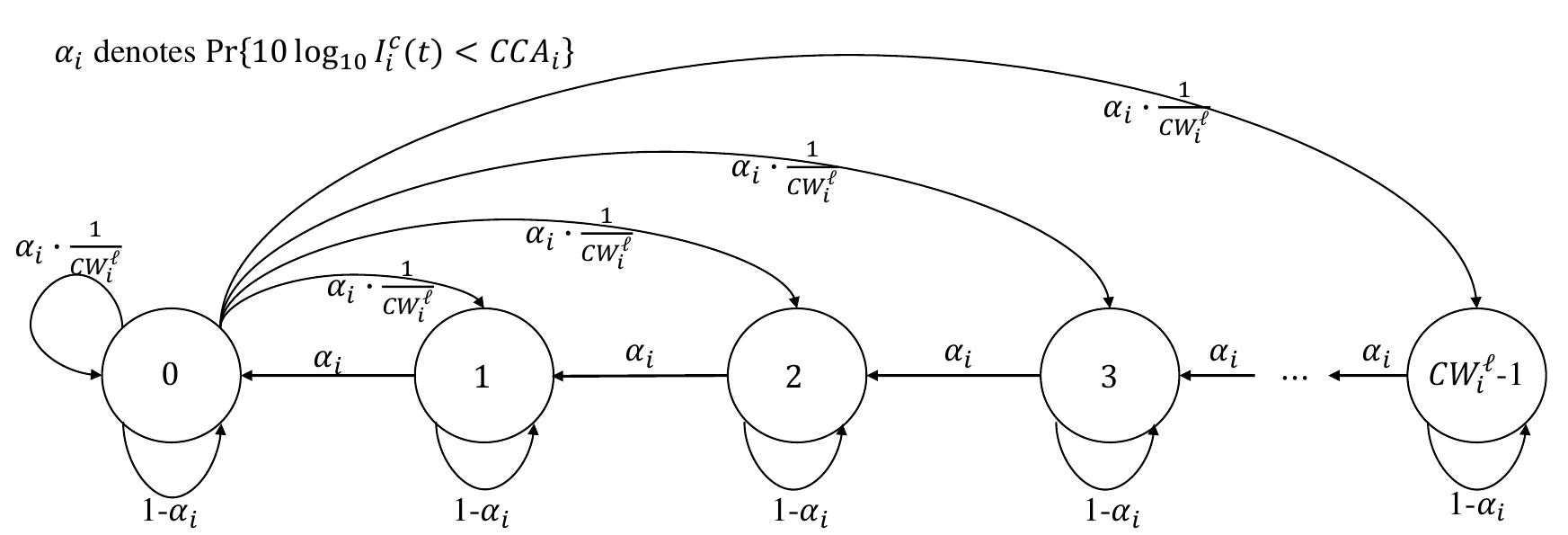}
    \caption{Backoff-state transition model for a typical AP $i$ in the $\ell$-th retransmission attempt with contention window $CW_i^{(\ell)}$.}
    \label{fig: state_transition}
\end{figure}

AP $i$ monitors the aggregate sub-7GHz noise-plus-interference power, denoted by $I_i^{\rm c}(t)$. The channel is regarded as idle only when
\begin{equation}
    10\log_{10} I_i^{\rm c}(t) < CCA_i .
\end{equation}
$I_i^{\rm c}(t)$ is maintained as an event-driven AP-side sensing variable. It is initialized by the noise power $N_0$ and is increased by the received sub-7GHz control-channel power from APs currently occupying the control channel
\begin{equation}
    I_i^{\rm c}(t)=N_0+\sum_{k\in\mathcal{N}_1(t),\,k\neq i}P_{k\to i}^{\rm c},
\end{equation}
where $\mathcal{N}_1(t)$ denotes the set of neighboring APs that may interfere with AP $i$ on the sub-7GHz control channel at time $t$, and
\begin{equation}
    P_{k\to i}^{\rm c}
    =P_{\rm tx}\left(\frac{c}{4\pi f_{\rm c}d_{k,i}}\right)^2,
\end{equation}
where $P_{\rm tx}$ is the transmit power of the AP, $c$ is the speed of light, $f_{\rm c}=6$ GHz, and $d_{k,i}$ is the corresponding AP $k$--AP $i$ distance. If the sensed power exceeds the selected CCA threshold, the AP defers for one slot and senses again. Once an idle channel is observed, the AP waits for a DIFS interval and then starts its backoff countdown. The backoff counter is drawn uniformly from $\{0,1,\ldots,CW_i^{(\ell)}-1\}$. For the $\ell$-th retransmission attempt of the same packet, the contention window follows binary exponential backoff, which is given by
\begin{equation}
    CW_i^{(\ell)}=CW_i2^{\min\{\ell,K\}},
\end{equation}
where $K$ is the maximum cutoff phase. The counter decreases by one after each idle slot and is frozen whenever the control channel becomes busy. To make the effect of $CCA_i$ and $CW_i^{(\ell)}$ explicit, define the probability that AP $i$ observes an idle control-channel slot as
\begin{equation}
    \alpha_i=\Pr\left\{10\log_{10} I_i^{\rm c}(t)<CCA_i\right\}.
\end{equation}
Fig.~\ref{fig: state_transition} shows the corresponding backoff-state transition model for the $\ell$-th attempt. A transition to the next lower counter state occurs with probability $\alpha_i$, while the counter remains frozen with probability $1-\alpha_i$. After the counter reaches zero and a new packet attempt is initiated, the next backoff state is selected according to the uniform distribution over the window $CW_i^{(\ell)}$.
Based on this abstraction, the average number of slots consumed before channel access includes the mean countdown value and one additional idle slot for the access attempt. Thus, the average number of backoff slots consumed before channel access in the $\ell$-th attempt can be obtained as \cite{dai2013unified}
\begin{equation}
    \tau_i^{(\ell)}=\frac{1}{\alpha_i} \cdot \frac{CW_i^{(\ell)}+1}{2}.
\end{equation}
The corresponding average backoff waiting time is $\tau_i^{(\ell)}\tau_{\rm slot}$, where $\tau_{\rm slot}$ is the slot duration. This expression separates the two sources of access delay. The contention window determines the average number of countdown steps, whereas the CCA threshold affects the probability that a slot can actually be used for countdown. Increasing $CCA_i$ generally increases $\alpha_i$ and shortens the waiting time, but it may also expose the AP to stronger hidden interference. Reducing $CW_i^{(\ell)}$ also shortens the waiting time, but it increases the probability that multiple APs finish backoff simultaneously and collide in the RTS phase.

\subsubsection{STA selection and beam training}
When the backoff counter reaches zero, AP $i$ selects the associated STA with the smallest instantaneous mmWave interference level as the intended receiver. If the beam training interval has expired, the AP performs beam training before data transmission. The training overhead is proportional to the number of angular sectors that must be swept and is modeled as
\begin{equation}
    T_i^{\rm tr}=T_{\rm pilot}\left\lceil\frac{360^\circ}{\theta_i}\right\rceil .
\end{equation}
Therefore, a narrower beam provides higher directional gain but incurs a larger training overhead. This creates a non-trivial tradeoff between link budget improvement, interference suppression, and control overhead.

\subsubsection{RTS/CTS control exchange}
After the backoff procedure, AP $i$ transmits an RTS frame over the sub-7GHz interface. During the RTS duration, the RTS signal contributes to the sensed power at neighboring APs, thereby preventing those APs from treating the control channel as idle. In this packet-level model, we do not separately model PHY decoding errors of sub-7GHz RTS/CTS control frames. Instead, RTS/CTS is used as a control-plane reservation procedure, while the CTS decision also performs a simplified mmWave admission check for the selected data mode.

Specifically, let $\mathcal{E}^{\rm adm}_{i,n}$ denote the event that the selected MCS can be supported at STA $n$ under the worst mmWave noise-plus-interference observed during the CTS-decision interval. If $\mathcal{E}^{\rm adm}_{i,n}$ does not hold, the CTS response is regarded as unsuccessful, AP $i$ waits for a CTS timeout, and the packet enters retransmission. If $\mathcal{E}^{\rm adm}_{i,n}$ holds, the receiver replies with CTS after SIFS, and the successful RTS/CTS handshake reserves the forthcoming mmWave data transmission opportunity. Thus, CTS failure in this model should be interpreted as mmWave admission rejection rather than as a detailed sub-7GHz control-frame decoding failure.

\subsubsection{Directional mmWave data transmission}
After a successful CTS, AP $i$ transmits the data payload on the mmWave link using beamwidth $\theta_i$. Let $P_{i,n}^{\rm mmW}$ denote the received mmWave power from AP $i$ to its intended STA $n\in\mathcal{S}_i$ before antenna directivity is applied. The simulator uses free-space path loss at 60 GHz,
\begin{equation}
    P_{i,n}^{\rm mmW}=P_{\rm tx}
    \left(\frac{c}{4\pi f_{\rm mm}d_{i,n}}\right)^2 ,
\end{equation}
where $f_{\rm mm}=60$ GHz, and $d_{i,n}$ is the corresponding AP--STA distance. The transmit antenna is represented by a sector model. For interference calculation, let $P_{k\to i,n}^{\rm mmW}$ denote the received power from an interfering AP $k$ to STA $n\in\mathcal{S}_i$ before antenna directivity is applied. If STA $n$ lies inside the main lobe of AP $k$, the interference gain is $G_{\rm m}(\theta_k)$; otherwise the sidelobe gain $G_{\rm s}$ is used. Hence, a concurrent transmission from AP $k$ contributes
\begin{equation}
    Q_{k\to i,n}(t)=G_{k\to i,n}(\theta_k)P_{k\to i,n}^{\rm mmW}
\end{equation}
to STA $n\in\mathcal{S}_i$, where $G_{k\to i,n}(\theta_k)\in\{G_{\rm m}(\theta_k),G_{\rm s}\}$ depends on the angular separation between the intended beam direction of AP $k$ and the victim receiver.

Let $\mathcal{N}_2(t)$ denote the set of APs that are transmitting on the mmWave band at time $t$. The noise-plus-interference power at STA $n\in\mathcal{S}_i$ is
\begin{equation}
    I_{i,n}^{\rm d}(t)=N_0+
    \sum_{k\in\mathcal{N}_2(t),\,k\ne i}Q_{k\to i,n}(t),
\end{equation}
where $N_0=10^{(-169+10\log_{10}(B_{\rm mm}))/10}$ mW is the thermal noise power over the mmWave bandwidth $B_{\rm mm}$ in Hz, and $-169$ dBm/Hz is the effective noise spectral density including the receiver noise figure. To separate admission control from data decoding, let $\mathcal{T}_i^{\rm adm}$ denote the CTS-decision interval and $\mathcal{T}_i^{\rm data}$ denote the subsequent mmWave payload interval. Since interfering transmissions may start or end during either interval, the conservative SINR over an interval $\mathcal{T}$ is
\begin{equation}
    \gamma_{i,n}=
    \frac{G_{\rm m}(\theta_i)P_{i,n}^{\rm mmW}}
    {\max_{t\in\mathcal{T}} I_{i,n}^{\rm d}(t)}.
\end{equation}
where the denominator includes both thermal noise and concurrent directional mmWave interference. The STA is modeled with omnidirectional reception, so no additional receive beamforming gain is applied at the receiver side. The selected PHY rate $R_i$ of AP $i$ is obtained from the IEEE 802.11ay MCS table \cite{ieee80211ay}\footnote{Here, we adopt the MCS table defined in the IEEE 802.11ay standard, as it is the latest available standard for mmWave WLANs and no IEEE P802.11bq standard is currently available.}. Specifically, the AP first estimates the instantaneous SINR before transmission, determines the highest supportable MCS, and then backs off from it according to the reservation margin $M_i$. The corresponding data duration is
\begin{equation}
    T_i^{\rm data}=\frac{L_{\rm p}+L_{\rm MAC}}{R_i},
\end{equation}
where $L_{\rm p}$ is the payload length, $L_{\rm MAC}$ is the MAC header length, and $R_i$ is the selected PHY rate. Let $\Gamma_{m_i}$ be the SINR threshold of the selected MCS. The admission event is $\mathcal{E}^{\rm adm}_{i,n}=\{\gamma_{i,n}(\mathcal{T}_i^{\rm adm})\geq\Gamma_{m_i}\}$, while the data-success event is $\mathcal{E}^{\rm data}_{i,n}=\{\gamma_{i,n}(\mathcal{T}_i^{\rm data})\geq\Gamma_{m_i}\}$. A packet is successfully delivered only when both events hold; otherwise, AP $i$ invokes the retransmission procedure.


\subsection{PHY/MAC Configuration and Action Space}

Assume that each AP $i$ autonomously selects a PHY/MAC configuration from a finite candidate space,
\begin{equation}
    \mathbf{x}_i=(CW_i,CCA_i,\theta_i,M_i)\in
    \mathcal{CW}\times\mathcal{CCA}\times\Theta\times\mathcal{M},
\end{equation}
where $\mathcal{CW}$, $\mathcal{CCA}$, $\Theta$, and $\mathcal{M}$ denote the candidate sets for the minimum contention window, the CCA threshold, the mmWave beamwidth for data transmission, and the MCS reservation margin, respectively. The concrete numerical values of these sets are part of the simulation setup and are reported in the simulation section.

The four decision variables influence different parts of the access and transmission pipeline. The minimum contention window $CW_i$ controls the initial random backoff range. A smaller $CW_i$ shortens the expected access delay and may improve channel utilization under light contention, but it also increases the probability that multiple APs finish backoff simultaneously and collide in the control phase. A larger $CW_i$ is more conservative and can reduce RTS contention, at the cost of longer idle waiting time.

The CCA threshold $CCA_i$ determines how sensitive AP $i$ is to sub-7GHz control-channel energy. A low threshold makes the AP defer more frequently, which protects ongoing transmissions but may excessively suppress spatial reuse. A high threshold allows the AP to access the channel in the presence of stronger sensed energy, thereby improving concurrency, but it can also aggravate hidden-terminal effects and increase failed RTS/CTS or data transmissions. Thus, $CCA_i$ directly governs the tradeoff between contention avoidance and spatial reuse.

The beamwidth $\theta_i$ shapes the mmWave data-plane interference pattern. Narrow beams provide higher main-lobe gain and lower interference leakage toward unintended receivers, which improves SINR and enables higher MCS choices. At the same time, narrow beams require more beam-training overhead because more angular sectors must be scanned. Wider beams reduce training overhead and are more tolerant to alignment uncertainty, but they provide lower directional gain and cause stronger interference to neighboring links.

The MCS reservation margin $M_i$ controls the conservativeness of link adaptation. Let $\Gamma_m$ be the required SINR threshold of MCS index $m\in\mathcal{Q}$. The highest supportable MCS is first obtained as
\begin{equation}
    m_i^{\max}=\max\{m\in\mathcal{Q}: \Gamma_m\leq \gamma_{i,n}\}.
\end{equation}
AP $i$ then applies the reservation margin by selecting
\begin{equation}
    m_i=\max\{m_{\min},m_i^{\max}-M_i\},
\end{equation}
where $m_{\min}$ is the lowest MCS index, and the PHY rate is set as $R_i=R(m_i)$. Thus, a small $M_i$ allows AP $i$ to use a more aggressive PHY rate when the estimated SINR is high, while a larger $M_i$ backs off from the highest supportable MCS to reserve additional SINR headroom against interference fluctuations. The optimal operating point therefore depends on both the local channel condition and the contention behavior of surrounding APs. These coupled effects motivate a learning-based configuration method that can optimize the joint action $\mathbf{x}_i$ from online throughput feedback.

\subsection{Throughput Reward and Optimization Objective}
The network-level performance metric considered in this paper is the aggregate throughput under distributed contention and time-varying interference. Let $s$ index the learning intervals, and let $C_i(t)$ denote the number of successfully delivered packets by AP $i$ up to time $t$. Over a learning interval $[t_s,t_{s+1})$, the short-term throughput of AP $i$ is
\begin{equation}
    r_i(t_s)=\frac{L_{\rm p}\left[C_i(t_{s+1})-C_i(t_s)\right]}
    {t_{s+1}-t_s}.
\end{equation}
This reward naturally reflects all effects of the selected configuration, including backoff delay, RTS/CTS failures, beam-training overhead, MCS selection, directional interference, and packet retransmissions. If complete network information and centralized control were available, the ideal network-utility benchmark could be written as
\begin{equation}
\begin{aligned}
    &\max_{\{\mathbf{x}_i\}_{i\in\mathcal{N}}}
    \sum_{i\in\mathcal{N}}\mathbb{E}\left[r_i(t_s)\right], \\
    &\text{s.t.} \quad
    \mathbf{x}_i\in\mathcal{CW}\times\mathcal{CCA}\times\Theta\times\mathcal{M},
    \quad \forall i\in\mathcal{N}.
\end{aligned}
    \label{eq:global_opt}
\end{equation}
Eq.~\eqref{eq:global_opt} is used as a system-level objective for evaluation rather than as the exact problem solved by each AP. In the considered distributed setting, AP $i$ does not observe neighboring APs' actions, traffic states, or interference maps. It only has access to its local learning history
\begin{equation}
    o_i(t)=\{(\mathbf{x}_i(t_0),r_i(t_0)):t_0<t\},
\end{equation}
and selects its next configuration through a local policy
\begin{equation}
    \mathbf{x}_i(t)=\pi_i(o_i(t)).
\end{equation}
Maximizing the local reward of each AP is not generally equivalent to solving \eqref{eq:global_opt}, because one AP's action changes the contention probability and directional interference experienced by other APs. Moreover, when multiple APs adapt simultaneously, the reward distribution observed by any single AP is non-stationary. Therefore, the proposed method should be interpreted as decentralized adaptive learning that seeks a high-throughput operating point using only local throughput feedback. The action space remains combinatorial: even for a single AP, a naive exhaustive learning approach must evaluate $(|\mathcal{CW}|\cdot|\mathcal{CCA}|\cdot|\Theta|\cdot|\mathcal{M}|)$ possible configurations. We therefore formulate the online configuration problem as a distributed combinatorial multi-armed bandit problem under local observations.

\section{Proposed Algorithm and Performance Optimization}\label{sec:Algorithm and Optimization}

The optimization variables introduced in the previous section jointly determine the access aggressiveness, the control-channel reservation behavior, the directional interference pattern, and the selected mmWave PHY rate. A direct optimization over the Cartesian product $\mathcal{CW}\times\mathcal{CCA}\times\Theta\times\mathcal{M}$ is conceptually simple, but it treats every complete PHY/MAC configuration as an unrelated arm. Such a flat bandit formulation is inefficient because the number of arms grows multiplicatively with the number of parameters, and it ignores the fact that each configuration is composed of interpretable elementary choices.

To exploit this structure, each AP is equipped with a CSAR-inspired multi-group combinatorial MAB agent \cite{chen2014combinatorial, rejwan2020top}. The method borrows the high-level successive accept-reject principle and Hadamard-guided probing idea from existing combinatorial bandit algorithms, but adapts them to the IEEE P802.11bq setting by enforcing group-wise feasibility and by using empirical ranking scores under non-additive packet-level throughput rewards. Therefore, the theoretical guarantees of the original CSAR formulations are not claimed here. The action is decomposed into $G=4$ parameter groups, corresponding to the contention window, CCA threshold, beamwidth, and MCS reservation margin. Let $\mathcal{V}_g$ be the candidate set of group $g \in G$, and let $V_g=|\mathcal{V}_g|$. A feasible compound action selects exactly one elementary arm from each group,
\begin{equation}
    \mathbf{x}=(v_1,v_2,\ldots,v_G),\quad v_g\in\mathcal{V}_g .
\end{equation}
The reward of this compound action is the normalized short-term throughput measured by the AP after the configuration is applied for one learning interval. This reward is stochastic because it depends on random backoff evolution, STA selection, beam-training overhead, packet-level SINR fluctuations, and the concurrent decisions of neighboring APs. Each AP does not require a parametric model of these effects; it only observes the realized throughput.

The central idea of the proposed variant is to avoid learning all $\prod_{g=1}^{G}V_g$ compound arms independently. Instead, it uses Hadamard-guided feasible exploration to generate diverse compound configurations and then computes empirical elementary-arm scores for group-wise pruning. These scores are used as ranking proxies rather than unbiased least-squares estimates of an additive reward model. This design preserves the feasibility of the IEEE P802.11bq configuration, since one candidate is always retained for every group, while reducing the number of exploratory trials required before each AP concentrates on high-throughput configurations.

\subsection{Elementary-Arm Indexing and Active Sets}
CSAR first converts the grouped action space into a single indexed list of elementary arms. Let
\begin{equation}
    V=\sum_{g=1}^{G}V_g
\end{equation}
be the total number of elementary arms. The elementary arms are indexed by concatenating the groups. The global index set corresponding to group $g$ is
\begin{equation}
    \mathcal{I}_g=\left\{1+\sum_{h<g}V_h,\ldots,\sum_{h\leq g}V_h\right\}.
\end{equation}
For example, all contention-window candidates occupy one contiguous block of indices, all CCA candidates occupy another block, and so on. This indexing enables the algorithm to estimate a score for every elementary arm while still enforcing the constraint that exactly one element must be chosen from each group.

For each group, CSAR maintains an active set $\mathcal{A}_g\subseteq\mathcal{I}_g$. Initially, $\mathcal{A}_g=\mathcal{I}_g$, meaning that all candidates are eligible for selection. At the end of each phase, one low-scoring candidate is removed from every group that still contains more than one active arm. Therefore, the active sets gradually shrink from broad exploration to a compact set of surviving configurations.

\subsection{Hadamard-Guided Feasible Exploration}
A key difficulty in the considered grouped configuration problem is that the agent must obtain informative observations about multiple elementary arms from a small number of compound-action trials while still executing only feasible IEEE P802.11bq configurations. Following the Hadamard probing idea used in full-bandit combinatorial learning, the proposed variant constructs a Hadamard matrix $\mathbf{H}\in\{-1,+1\}^{L\times L}$, where
\begin{equation}
    L=2^{\lceil \log_2 V\rceil}.
\end{equation}
Here, $L$ is the smallest power of two no smaller than the total number of elementary arms. The first $V$ columns of $\mathbf{H}$ are assigned to the $V$ elementary arms, while the remaining columns, if any, are padding columns and are ignored when scores are computed.

Each phase consists of one complete scan of the $L$ Hadamard rows. For row $k$, CSAR identifies the elementary arms whose corresponding Hadamard entries are positive,
\begin{equation}
    \mathcal{P}_k=\{a\in\{1,\ldots,V\}:H_{k,a}=+1\}.
\end{equation}
The row itself is not a directly executable action, because it may mark zero, one, or multiple active arms in the same parameter group, whereas a feasible IEEE P802.11bq configuration must select exactly one candidate from each group. CSAR therefore maps the row to a feasible compound action group by group. For group $g$, if $\mathcal{P}_k\cap\mathcal{A}_g$ is nonempty, the agent selects one arm from this intersection. If the intersection is empty, it samples one arm from $\mathcal{A}_g$ as a fallback. This mapping guarantees that every trial specifies one valid contention window, one CCA threshold, one beamwidth, and one MCS reservation margin. After the selected elementary arms are mapped back to physical parameters, the AP applies the resulting feasible configuration for one learning interval and observes a throughput reward $y_k$. The corresponding Hadamard row is therefore used to guide exploration in the action space. By scanning all rows, CSAR obtains a structured set of diverse feasible configurations while maintaining the grouped feasibility constraint.

\subsection{Empirical Main-Effect Scoring}
At the end of a phase, the $L$ rewards are collected into $\mathbf{y}=[y_1,\ldots,y_L]^{\rm T}$. Because the actually executed compound action is obtained after group-wise feasible mapping and possible fallback sampling, the original Hadamard matrix is not the exact design matrix of the executed actions. We therefore do not interpret the following score as a strict least-squares estimator or as the elementary-arm estimator used in the original CSAR analyses. Instead, the proposed variant computes a Hadamard-guided empirical main-effect score
\begin{equation}
    \widehat{\boldsymbol{\omega}}
    =\frac{1}{L}\mathbf{H}^{\rm T}\mathbf{y}.
    \label{eq:hadamard_est}
\end{equation}
Only the first $V$ entries of $\widehat{\boldsymbol{\omega}}$ are used, because only these entries correspond to actual elementary arms. A larger score indicates that Hadamard rows favoring the corresponding elementary arm tend to produce higher rewards after feasible mapping within the current phase. Thus, $\widehat{\boldsymbol{\omega}}$ is used as a ranking proxy for successive rejection, not as an unbiased estimate of an additive WLAN throughput model.

\subsection{Group-Wise Successive Rejection}
After score estimation, the proposed variant applies successive pruning separately within each parameter group. For every group that still has more than one active candidate, the lowest-scoring elementary arm is removed:
\begin{equation}
    a_g^{-}=\arg\min_{a\in\mathcal{A}_g}\widehat{\omega}_a,\qquad
    \mathcal{A}_g\leftarrow \mathcal{A}_g\setminus\{a_g^{-}\}.
\end{equation}
Note that it discards only one candidate per group per phase. It also preserves the group structure of the action space: after rejection, each future compound action still contains one valid CW value, one valid CCA threshold, one valid beamwidth, and one valid MCS margin. The algorithm then resets the phase rewards and begins another Hadamard-guided scan over the reduced active sets.

This phase-wise rejection process continues until one active arm remains in each group. The surviving arms form the learned PHY/MAC configuration. If the environment continues to run after convergence, the AP repeatedly applies the configuration generated from these surviving active sets. The sample cost of one phase is $L$, with $L\leq 2V$. Since at most one candidate is removed from every non-singleton group in one phase, the number of rejection phases is bounded by $\max_g(V_g-1)$. Hence, the exploration effort scales with the total number of elementary arms and the number of phases, rather than with the full Cartesian-product action space.

\subsection{Theoretical Performance Analysis}
\textit{Feasibility and finite termination:} At every exploration trial, the mapping from a Hadamard row to a compound action selects exactly one active candidate from each group. Hence, for all phases and all learning intervals,
\begin{equation}
    \mathbf{x}(t_s)\in\mathcal{V}_1\times\cdots\times\mathcal{V}_G .
\end{equation}
Moreover, each completed phase removes at most one candidate from every group that still contains more than one active arm and never removes the last remaining candidate of any group. Therefore, the algorithm terminates after at most
$\max_{g\in\{1,\ldots,G\}}(V_g-1)$
rejection phases, with one feasible candidate left in each parameter group.

\textit{Exploration complexity:} Each phase scans $L=2^{\lceil\log_2 V\rceil}$ Hadamard rows, where $V=\sum_{g=1}^{G}V_g$. Since $L<2V$ for non-power-of-two $V$, the total number of learning intervals before termination is bounded by
\begin{equation}
    N_{\rm expl}\leq L \max_{g\in\{1,\ldots,G\}}(V_g-1)<2V\max_{g\in\{1,\ldots,G\}}(V_g-1).
\end{equation}
This is linear in the total number of elementary arms up to the number of rejection phases. In contrast, a flat Cartesian-arm bandit must consider $N_{\rm cart}=\prod_{g=1}^{G}V_g$ complete configurations. 


\begin{algorithm}[t]
\caption{CSAR-Inspired Group-Wise Feasible Configuration Learning for One AP}
\label{alg:csar}
\begin{algorithmic}[1]
\State Index all elementary arms and initialize $\mathcal{A}_g\gets\mathcal{I}_g$, $g=1,\ldots,G$
\State Construct $\mathbf{H}\in\{-1,+1\}^{L\times L}$ with $L=2^{\lceil\log_2 V\rceil}$
\While{at least one group has more than one active arm}
    \For{$k=1,\ldots,L$}
        \State $\mathcal{P}_k\gets\{a\in\{1,\ldots,V\}:H_{k,a}=+1\}$
        \For{$g=1,\ldots,G$}
            \State Select $a_g$ from $\mathcal{P}_k\cap\mathcal{A}_g$ if nonempty;
            \State otherwise, sample $a_g$ from $\mathcal{A}_g$
        \EndFor
        \State Map $(a_1,\ldots,a_G)$ to $(CW,CCA,\theta,M)$ and apply this feasible configuration for one learning interval
        \State Observe the normalized short-term throughput reward $y_k$
    \EndFor
    \State Compute empirical scores using $\widehat{\boldsymbol{\omega}}=L^{-1}\mathbf{H}^{\rm T}\mathbf{y}$
    \For{$g=1,\ldots,G$}
        \If{$|\mathcal{A}_g|>1$}
            \State Remove $\arg\min_{a\in\mathcal{A}_g}\widehat{\omega}_a$ from $\mathcal{A}_g$
        \EndIf
    \EndFor
\EndWhile
\State \Return the configuration composed of the remaining arms
\end{algorithmic}
\end{algorithm}

\subsection{Online Operation in the IEEE P802.11bq Access Process}
The proposed learner operates at each AP as an asynchronous distributed controller. At the beginning of a learning interval, the AP obtains a PHY/MAC configuration from CSAR and applies it to the CSMA/CA and mmWave transmission procedures. During this interval, the AP follows the complete packet service process described in the system model, including carrier sensing, backoff countdown, RTS/CTS exchange, optional beam training, mmWave payload transmission, ACK reception, and retransmission after failure.

At the end of the interval, the AP computes its local short-term throughput and feeds the normalized reward back to CSAR. The learner stores this reward together with the Hadamard row that guided the feasible configuration. If the current phase has not yet scanned all $L$ rows, the learner advances to the next row and produces the next feasible configuration. Once a phase is complete, it computes the empirical elementary-arm scores using \eqref{eq:hadamard_est}, removes the weakest active candidate in each non-singleton group, resets the phase buffer, and starts the next phase.

This closed-loop operation optimizes the same throughput metric observed during packet service. It does not require the actions, traffic states, or interference maps of neighboring APs; their effects are implicitly captured by the local reward. Consequently, the proposed Algorithm \ref{alg:csar} can adapt the four system parameters toward a balanced operating point: it can reduce overly aggressive contention settings when they cause RTS/CTS failures, favor narrower beams when the SINR gain outweighs training overhead, or select a more conservative MCS margin when interference fluctuations dominate packet reliability.

\begin{table}[t]
	\centering
    \vspace{-0.5cm}
    \renewcommand\arraystretch{1.25}
	\caption{Simulation parameter configuration.} 
    \label{table: parameter}
	\setlength{\tabcolsep}{2.6mm}
	\begin{tabular}{p{0.41\columnwidth}p{0.47\columnwidth}}
        \toprule
		Parameter & Value \\
        \midrule
        Number of APs, $N$ & $\{2,4,6,8,10,12\}$ \\
        Number of STAs per AP & Uniform in $[4,10]$ \\
        AP--STA range offset & $[2,10]$ m \\
        Channel bandwidth $B$ & $80$ MHz \\
        Control/data plane frequencies & $6$ GHz / $60$ GHz \\
        Transmit power $P_{\rm tx}$ & $23$ dBm \\
        Noise spectral density & $-169$ dBm/Hz \\
        Payload length $L_{\rm p}$ & $240000$ bits \\
        MAC header length $L_{\rm MAC}$ & $288$ bits \\
        Slot/DIFS/SIFS/PIFS & $9/34/16/25~\mu$s \\
        RTS/CTS/CTS timeout & $10/8/41~\mu$s \\
        ACK duration & $40~\mu$s \\
        Beam-training pilot duration & $160~\mu$s \\
        Beam-training interval & $25$ ms \\
        Maximum backoff stage $K$ & $6$ \\
        CW candidates $\mathcal{CW}$ & $\{4,8,16,32,64,128\}$ \\
        CCA candidates $\mathcal{CCA}$ & $\{-82,-50,-30,-10,0,10\}$ dBm \\
        Beamwidth candidates $\Theta$ & $\{14.5,28.5,43.5,59.5\}^{\circ}$ \\
        Main-lobe gains $G_{\rm m}$ & $\{21.7,11.1,7.3,5.4\}$ dB \\
        Side-lobe gain $G_{\rm s}$ & $-8.86$ dB \\
        MCS reservation set $\mathcal{M}$ & $\{0,1,2,3,4,5,6\}$ \\
        CSAR update interval & $10$ ms \\
        TS-MAB update interval & $3$ ms \\
        Simulation time per episode & $5$ s \\
        Number of episodes & $10$ \\
        \bottomrule
	\end{tabular}
\end{table}

\section{Simulation Results} \label{sec:Simulation}
In this section, we evaluate the proposed CSAR-based configuration learning scheme through packet-level simulations. The simulation follows the multi-AP IEEE P802.11bq model described in Section~\ref{sec:System Model And Preliminary Analysis}, where the sub-7GHz interface is used for carrier sensing and RTS/CTS control exchange, while the mmWave interface is used for directional data transmission. Unless otherwise specified, each data point is averaged over 10 independent episodes with different random seeds. The main performance metrics are the aggregate network throughput, the per-AP throughput, and the collision rate, where the collision rate is defined as the average number of collision events per second. The parameter settings are summarized in Table~\ref{table: parameter}.

Simulation setup details are as follows. For each AP-density scenario, APs are placed at fixed coordinates specified by the scenario file. Each AP serves a uniformly drawn number of associated STAs in $[4,10]$; when STA coordinates are not explicitly specified, their horizontal and vertical offsets from the serving AP are independently drawn from $[2,10]$ m with random signs, so the AP--STA association is fixed by construction. All APs operate under saturated traffic. For every packet, all schemes use the same local scheduler, which selects the associated STA with the smallest instantaneous mmWave interference. The sub-7GHz control plane uses 6 GHz free-space AP-to-AP sensing/interference and CCA-based deferral; no separate PHY decoding error is introduced for RTS/CTS/ACK, and ACK timing is counted after the mmWave payload. Failed packets are retransmitted with the backoff stage capped at $K=6$, and beam-training timers start from zero and follow the $25$ ms interval in Table~\ref{table: parameter}.

\begin{figure*}[t]
\vspace{-1cm}
\centering
\subfloat[]{\includegraphics[width=3.4in,height=2.4in]{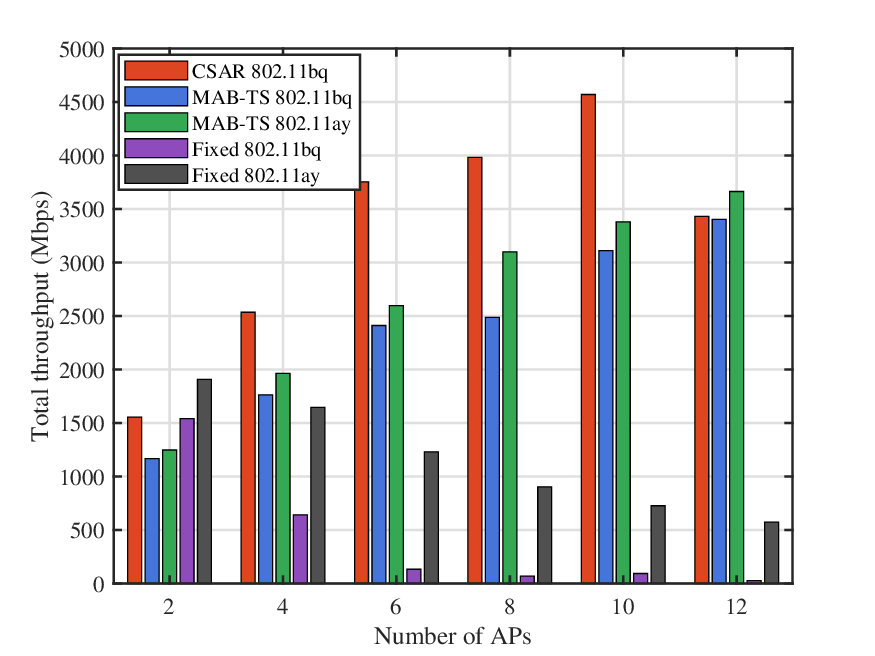}
	\label{fig: network_throughput_vs_ap}} \quad
\subfloat[]{\includegraphics[width=3.4in,height=2.4in]{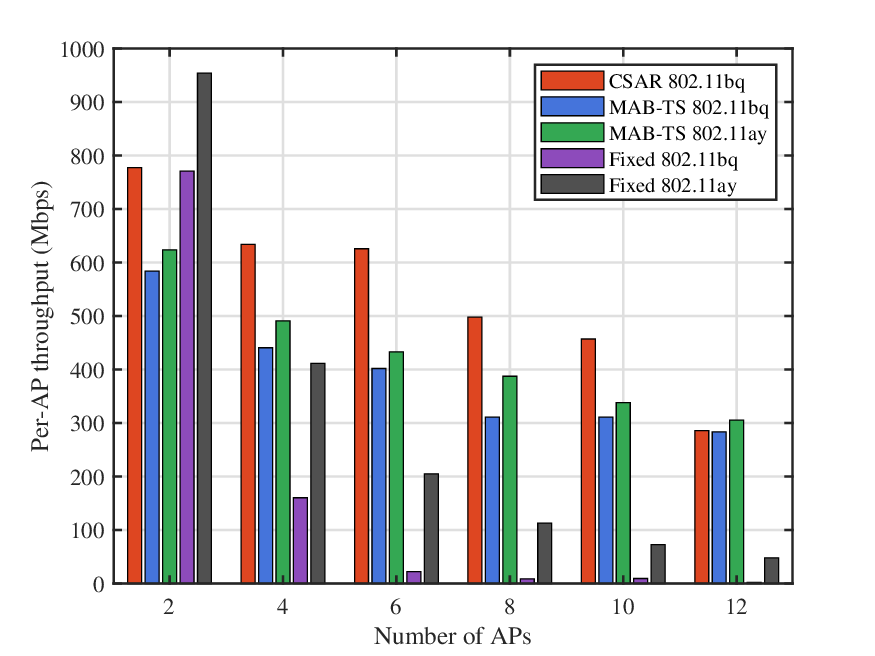}
        \label{fig: per_ap_throughput_vs_ap}}
    \caption{(a) Throughput performance of the proposed CSAR-based learning algorithm compared with baseline configurations. (b) Per-AP throughput comparison among the proposed CSAR scheme and the considered baselines.}
\label{fig: throughput_vs_ap}
\end{figure*}

We compare the proposed CSAR-based IEEE P802.11bq scheme with four baselines. The first baseline is an IEEE P802.11bq network optimized by Thompson sampling (MAB-TS), which treats each complete PHY/MAC configuration as an arm. The second baseline is a pure-mmWave IEEE 802.11ay-like network optimized by MAB-TS. The last two baselines are fixed-configuration versions of IEEE P802.11bq and pure-mmWave operation, respectively. This comparison allows us to examine not only the learning algorithm itself, but also the benefit of jointly coordinating sub-7GHz contention control and mmWave data transmission.

For the fixed baseline, $(CW,\mathrm{CCA},\theta,M)=(4,-82~\mathrm{dBm},14.5^\circ,0)$, where $M$ denotes the MCS reservation margin. The fixed pure-mmWave baseline uses $CW=4$, $\theta=14.5^\circ$, and $M=0$ under the pure-mmWave access model. In the ablation study, MAC-only CSAR fixes $(\theta,M)=(14.5^\circ,0)$, while PHY-only CSAR fixes $(CW,\mathrm{CCA})=(16,-82~\mathrm{dBm})$. The TS baselines use the full Cartesian configuration space as arms, zero initial Gaussian means, unit initial precision, an initial arm scan, reward normalization by $10$, and a $3$ ms update interval. This update interval is selected after tuning the TS baseline; using the same $10$ ms interval as CSAR substantially degrades TS, so the reported comparison uses the stronger tuned baseline.

Table~\ref{table: parameter} lists both the physical-layer parameters and the MAC-layer timing parameters used in the simulations. The CSAR action space contains four groups: the contention window, the CCA threshold, the mmWave beamwidth, and the MCS reservation offset. Hence, a complete configuration is formed by selecting one candidate from each group. This grouped structure is exploited by CSAR, while the MAB-TS baseline learns over complete configurations directly.

\begin{figure}[t]
    \centering
    \includegraphics[width=0.45\textwidth]{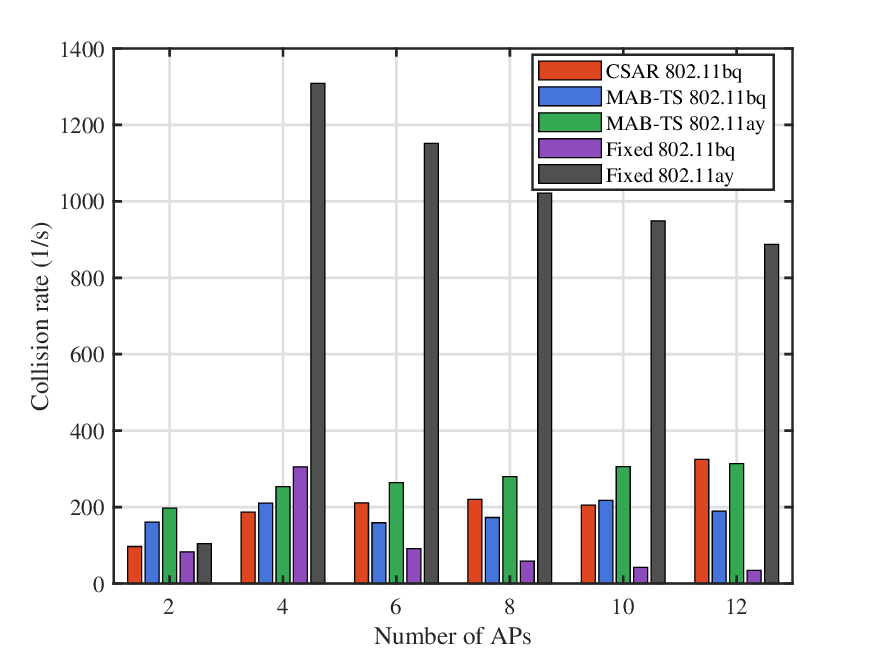}
    \caption{Collision performance of the proposed CSAR-based learning algorithm compared with baseline configurations.}
    \label{fig: collision_vs_ap}
\end{figure}

Fig.~\ref{fig: throughput_vs_ap} shows the throughput performance under different AP densities. In Fig.~\ref{fig: network_throughput_vs_ap}, the proposed CSAR-based IEEE P802.11bq scheme achieves a clear throughput gain over the MAB-TS-based IEEE P802.11bq baseline for most AP densities. For example, when the number of APs increases from 4 to 10, CSAR improves the aggregate throughput from $2535.16$ Mbps to $4570.00$ Mbps, while the corresponding MAB-TS baseline increases from $1762.69$ Mbps to $3110.28$ Mbps. This indicates that the grouped combinatorial learning structure is useful in dense deployments, because it avoids spending excessive samples on complete configurations that differ only in one parameter component.

The fixed IEEE P802.11bq baseline performs well only in the sparse case. As the number of APs increases, its throughput rapidly collapses, which confirms that a single static configuration cannot handle the change from a lightly loaded network to a dense multi-AP network. In contrast, CSAR maintains high throughput by adapting the contention behavior and the mmWave transmission parameters jointly. Compared with the pure-mmWave MAB-TS baseline, CSAR achieves higher throughput from 2 to 10 APs, while the pure-mmWave baseline becomes slightly better at 12 APs. This observation is meaningful: the proposed IEEE P802.11bq design is not simply relying on the mmWave data rate, but on the interaction between robust sub-7GHz control and directional mmWave spatial reuse. When the network becomes extremely dense, the benefit of aggressive mmWave spatial reuse can still be competitive, which motivates the cross-layer parameter adaptation studied in this paper.

\begin{figure*}[t]
    \centering
    \vspace{-0.5cm}
    \includegraphics[width=0.9\textwidth]{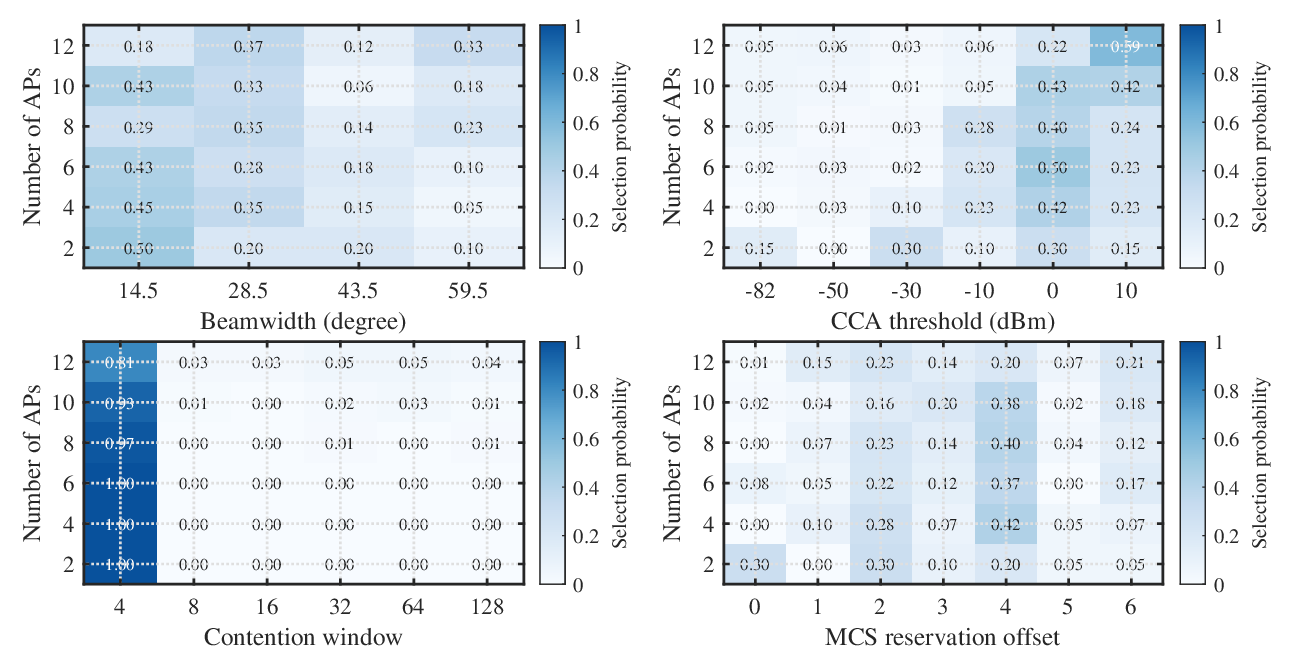}
    \caption{Parameter selection-probability heatmaps of the four CSAR-controlled parameters under different AP densities.}
    \label{fig: csar_parameter_heatmaps}
\end{figure*}

Fig.~\ref{fig: per_ap_throughput_vs_ap} further reports the per-AP throughput. For all schemes, the per-AP throughput decreases as more APs are deployed in the same area, because each AP experiences stronger contention and more directional interference. However, CSAR provides the best or near-best per-AP throughput among learning-based schemes in most cases. The gap is especially visible from 4 to 10 APs, where CSAR balances the increased spatial reuse opportunity with the increased collision risk. The per-AP result also explains why the total throughput does not increase indefinitely: after a certain density, adding more APs creates additional interference and control contention faster than it creates useful parallel transmission opportunities.

Fig.~\ref{fig: collision_vs_ap} compares the collision rate of different schemes. The fixed pure-mmWave baseline suffers from a very high collision rate in dense cases, exceeding $1000$ collisions/s for several AP densities. This is because pure-mmWave operation lacks the robust sub-7GHz control-plane coordination used by IEEE P802.11bq, and the fixed parameter setting cannot react to changes in interference and contention. The MAB-TS pure-mmWave baseline reduces the throughput degradation but still produces a relatively high collision rate as the network becomes denser.

The proposed CSAR scheme keeps the collision rate within a moderate range while maintaining high throughput. This is an important insight: the best throughput is not obtained by minimizing collisions alone. For instance, a very conservative configuration can reduce collisions but may also waste transmission opportunities. CSAR instead learns a balanced operating point where some level of spatial reuse is allowed, but excessive contention and packet failure are avoided. This behavior is consistent with the design of IEEE P802.11bq, where the sub-7GHz control channel should coordinate access without completely suppressing concurrent mmWave transmissions.

\begin{table}[t]
\centering
\caption{Throughput stabilization time comparison.}
\label{table: stabilization_time}
\begin{tabular}{|c|c|c|c|}
\hline
 number of APs &  CSAR &  MAB-TS &  CSAR gain (s) \\
\hline
       2 &  1.72 &   3.489 &        1.769 \\\hline
       4 &  1.66 &   3.687 &        2.027 \\\hline
       6 &  1.85 &   3.669 &        1.819 \\\hline
       8 &  1.96 &   3.783 &        1.823 \\\hline
      10 &  1.91 &   3.711 &        1.801 \\\hline
      12 &  2.21 &   3.648 &        1.438 \\
\hline
\end{tabular}
\end{table}

Table~\ref{table: stabilization_time} further compares the learning efficiency of CSAR and MAB-TS in terms of throughput stabilization time. This metric is distinct from the algorithm termination time of CSAR. With $V=23$ elementary arms, the Hadamard size is $L=32$; since the largest parameter group has $7$ candidates and the update interval is $10$~ms, completing all rejection phases requires up to $32\times 6\times 10$~ms $=1.92$~s.

The reported stabilization time is computed from the averaged throughput trace of each AP-density setting. Specifically, we first smooth the short-term throughput with a $0.5$~s rolling window, define the steady level as the mean throughput over the final $20\%$ of the trace, and record the first time at which the smoothed curve enters and then remains within $\pm5\%$ of this steady level. Under this definition, CSAR reaches a stable throughput earlier than the tuned Cartesian-arm MAB-TS baseline: its stabilization time is $1.66$--$2.21$~s, whereas MAB-TS requires about $3.49$--$3.78$~s. On average, CSAR shortens the stabilization time by about $1.78$~s, corresponding to an approximately $49\%$ reduction.

This faster throughput stabilization comes from the grouped combinatorial structure of CSAR. MAB-TS treats each complete PHY/MAC tuple as an independent arm, so configurations that differ only in one parameter must still be explored separately. In contrast, CSAR computes empirical scores for elementary arms and rejects inferior candidates within each parameter group. Therefore, useful information obtained from one compound configuration can accelerate the evaluation of other configurations sharing the same CW, CCA threshold, beamwidth, or MCS margin.

Fig.~\ref{fig: csar_parameter_heatmaps} and Table~\ref{table: csar_mean_parameters} provide insight into how CSAR adapts the four configuration groups. The beamwidth heatmap shows that narrow and moderate beams are frequently selected in sparse and medium-density cases, while wider beams become more likely when the network reaches 8 and 12 APs. This result reflects a tradeoff between antenna gain and beam-training overhead. Narrow beams provide stronger main-lobe gain and better interference suppression, but they also increase the cost of beam training. When the network becomes dense, CSAR sometimes favors wider beams to reduce training overhead and maintain service opportunities.

\begin{table}[t]
    \centering
    \renewcommand\arraystretch{1.2}
    \caption{Mean CSAR-selected parameters under different AP densities. The reported mean CW and MCS-margin values are rounded to the nearest integers.}
    \label{table: csar_mean_parameters}
    \setlength{\tabcolsep}{1.15mm}{
    \begin{tabular}{|c||c|c|c|c|c|c|}
        \hline
        \begin{tabular}{@{}c@{}}Mean selected\\parameter\end{tabular} & $2$ APs & $4$ APs & $6$ APs & $8$ APs & $10$ APs & $12$ APs \\
        \hline
        $\bar{\theta}$ & $27.6$ & $26$ & $28.28$ & $33.51$ & $28.96$ & $37.88$ \\
        \hline
        $\bar{CCA}$ & $-20.8$ & $-4.5$ & $-3.2$ & $-5.85$ & $-2.7$ & $-2.43$ \\
        \hline
        $\bar{CW}$ & $4$ & $4$ & $4$ & $6$ & $8$ & $14$ \\
        \hline
        $\bar{M}$ & $2$ & $3$ & $3$ & $3$ & $4$ & $3$ \\
        \hline
    \end{tabular}}
\end{table}

The CCA threshold distribution moves toward more aggressive values as the AP density increases. In particular, thresholds around $0$ dBm and $10$ dBm obtain high selection probabilities in dense deployments. This indicates that overly sensitive carrier sensing would unnecessarily defer transmissions in a dense indoor mmWave WLAN. CSAR therefore learns to relax the CCA threshold, allowing more spatial reuse on the mmWave data plane. At the same time, the contention-window heatmap shows that the smallest window remains dominant, but larger windows appear more frequently at high AP densities. This suggests that CSAR does not simply choose an aggressive access policy; instead, it combines a relaxed CCA threshold with occasional contention-window enlargement to control collision risk.

The MCS reservation offset is mainly concentrated around intermediate values. This means that CSAR avoids both overly aggressive rate selection and overly conservative rate reduction. A small offset may choose a high MCS that is vulnerable to interference fluctuations, while a large offset sacrifices PHY rate. The learned distribution therefore confirms that MCS reservation acts as a reliability margin for directional mmWave transmission. The averaged values in Table~\ref{table: csar_mean_parameters} summarize these trends: as the number of APs increases, the mean CCA threshold becomes less conservative, the mean contention window increases in dense cases, and the MCS reservation offset remains in a moderate range.

\begin{figure}[t]
    \centering
    \vspace{-0.5cm}
    \includegraphics[width=0.45\textwidth]{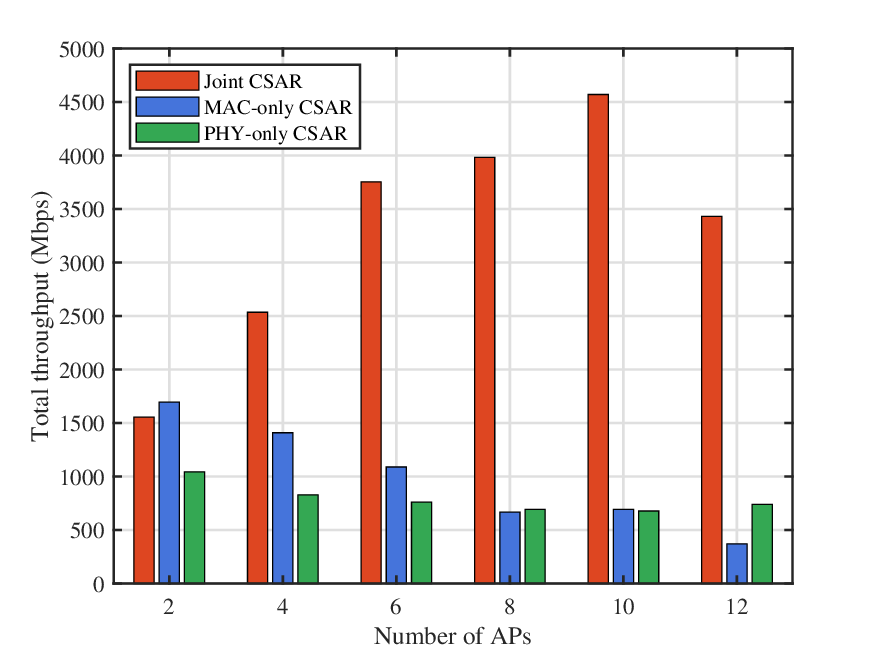}
    \caption{Total throughput comparison for different ablation settings.}
    \label{fig: ablation_total_throughput}
\end{figure}

Finally, Fig.~\ref{fig: ablation_total_throughput} evaluates the contribution of joint PHY/MAC optimization. The joint CSAR scheme is compared with two restricted variants: MAC-only CSAR, where the PHY-related parameters are fixed, and PHY-only CSAR, where the MAC-related parameters are fixed. The joint scheme achieves the highest throughput in almost all AP densities and becomes significantly better as the network becomes denser. This confirms that optimizing only one side of the system is insufficient. MAC-only adaptation can improve channel access, but it cannot adjust beamwidth or MCS robustness to directional interference. PHY-only adaptation can improve link reliability, but it cannot resolve contention and RTS/CTS collisions caused by inappropriate access parameters.

The ablation result also explains the effectiveness of the proposed combinatorial formulation. The four decision variables do not contribute independently; rather, their effects are coupled through the packet service process. A relaxed CCA threshold is useful only when the selected beamwidth and MCS margin can tolerate the resulting interference, and a small contention window is beneficial only when collisions remain manageable. By learning over grouped PHY/MAC parameters simultaneously, CSAR captures this coupling and produces a more robust configuration than either restricted variant.

\section{Conclusion and Discussion} \label{sec:Conclusion_discussion}
This paper investigated distributed throughput optimization for dense multi-AP IEEE P802.11bq WLANs. We first developed a packet-level system model that captures the coupling between sub-7GHz control-plane contention and directional mmWave data transmission. The model includes carrier sensing, random backoff, RTS/CTS exchange, beam-training overhead, SINR-based MCS selection, directional interference, and retransmission behavior. Based on this model, we formulated the PHY/MAC configuration problem as a multi-group combinatorial bandit, where each AP jointly selects the contention window, CCA threshold, beamwidth, and MCS reservation margin from finite candidate sets.

To solve this problem, we developed a CSAR-inspired group-wise feasible learning method that exploits the grouped structure of the configuration space. Instead of treating every complete parameter tuple as an independent arm, the method decomposes each configuration into elementary arms, computes empirical ranking scores through Hadamard-guided feasible exploration, and progressively prunes low-scoring candidates within each parameter group. This design should be viewed as an IEEE P802.11bq adaptation of existing combinatorial successive accept-reject ideas, rather than as a direct application of their additive-reward theoretical guarantees.

Simulation results showed that the proposed scheme improves the aggregate and per-AP throughput over the MAB-TS-based IEEE P802.11bq baseline for most AP densities, and prevents the severe performance degradation observed under fixed configurations. The collision-rate results further showed that high throughput is not achieved by simply minimizing collisions, but by learning a balanced operating point that allows useful mmWave spatial reuse while controlling excessive RTS/CTS failures and packet losses. The parameter-selection heatmaps revealed interpretable adaptation behavior: CSAR tends to relax the CCA threshold in dense deployments, occasionally enlarges the contention window to control contention, and chooses moderate MCS reservation offsets to balance rate and reliability. The ablation study confirmed that joint PHY/MAC optimization is essential, since MAC-only or PHY-only adaptation cannot fully capture the interaction between channel access, beamforming overhead, and directional interference.

Future work will extend the current framework toward more realistic deployment conditions, including mobility, blockage dynamics, heterogeneous traffic demands, and multi-channel operation. It is also promising to incorporate fairness-aware or latency-aware rewards so that IEEE P802.11bq networks can jointly optimize throughput, service regularity, and user-level quality of experience.

\bibliographystyle{IEEEtran}
\bibliography{IEEEabrv,related_work_references}
\balance


\end{document}